\documentstyle[12pt]{article}
\begin{document}

\title{
\flushright{
{\small UCRL-JC-115713 \\ Preprint}}
\flushleft
The Multifragmentation Freeze--Out Volume in Heavy Ion Collisions}

\author{
G.~Peilert, T.C.~Sangster, M.N.~Namboodiri, and H.C.~Britt
\\ \hspace{1cm} \\
Lawrence Livermore National Laboratory,
\\ Livermore, CA 94550
\\ USA}
\maketitle

{\bf Abstract:}

The reduced velocity correlation function for fragments from the
reaction Fe + Au at 100 A~MeV bombarding energy is investigated using
the dynamical--statistical approach QMD+SMM and compared to
experimental data to extract the Freeze--Out volume assuming
simultaneous multifragmentation.  It is shown that the data are
consistent with a Freeze--Out volume corresponding to $0.1 - 0.3$
times normal nuclear matter density.  The calculations show an
additional correlation due to recoil effects from a heavy third
fragment in an asymmetric break--up. This effect is present, but less
pronounced in the data.

\clearpage

\setcounter{page}{1}
A topic of great interest in nuclear physics is the multifragmentation
of heavy nuclei at moderate excitation energies.  Such systems are
typically formed in p-nucleus reactions in the GeV energy regime or in
intermediate energy heavy ion collisions. In the experiments of the
Purdue group \cite{perdue}, first attempts were made to relate the
inclusive results from p--nucleus fragmentation to the critical
exponents of a phase transition, analogous to the liquid gas
transition in condensed matter physics. Within the last few years it
has become possible to study such reactions in highly exclusive
experiments in order to extract the critical properties. Critical
properties can then be extracted investigating by, for example the
moments of the mass distribution as proposed by Campi \cite{campi}.

Such a procedure, however can not provide the physical quantities that
drive this phase transition. If one wants to extract quantities like a
critical temperature and density one has to rely on dynamical models.
Unfortunately there is presently no complete model available that
describes the process of thermally driven multifragmentation in heavy
ion collisions in a consistent approach. Current modeling
involves a two--step dynamical--statistical process beginning with a
pre equilibrium stage described in the earliest approaches by an
Intranuclear Cascade Model \cite{INC} and, more recently, by molecular
dynamics pictures such as the Quantum Molecular Dynamics model, QMD
\cite{QMD1,QMD2} or by single particle models of the VUU/BUU type
\cite{vuu}.  Following this stage there remains a distribution of
nucleons and complex fragments which can themselves be highly excited
and will undergo further statistical decay.  This statistical decay
process has been modeled in various codes as a sequential evaporation
\cite{blann,gemini,frdm}  or  as an explosive simultaneous multifragmentation
\cite{fai,bondorf,smm,gross,qsm}.  None of these models, however, includes
the dynamics of the fragmentation.

Recent investigations with the QMD model have shown that for central
collisions of heavy nuclei at high energies, a rapid
compression--decompression mechanism emerges in the initial stage of
the reaction and leads to a direct multifragmentation process, where a
highly excited heavy residue is no longer formed with high probability
\cite{QMD2,lynch93}.  These investigations suggest that the region to
search for the thermally driven multifragmentation in heavy ion
reactions is in central collisions at moderate bombarding energies
(E/A $\approx$ 100 A MeV) or in peripheral (or highly asymmetric)
reactions at higher energies (E/A $\approx$ 1 A GeV) . Under these
conditions the direct reaction leads to a highly excited source that
then breaks up into many IMF's.

Recently, attempts have been made  by Sangster et al., \cite{craig}
and Lacey at al. \cite{lacey} to extract the emission time scale of
the multifragmentation process by assuming a sequential decay
mechanism with a fixed Freeze-Out volume.  In
contrast to earlier work performed by Kim et al.
\cite{kim} using the Koonin--Pratt formalism \cite{kooninpratt} they
used  a the
classical three-body trajectory code {\it MENEKA \cite{meneka}} to
simulate the \underline{sequential} emission of IMF's from the  surface  of
a spherical source
characterized by a unique radius parameter. In this Letter we will
show, that the same data can be explained by assuming a
\underline{simultaneous} multifragmentation from a Freeze--Out volume
considerably
larger than the ground state dimension.

This will be done by using the QMD+SMM approach. In this case the
Hamilton equations of all nucleons are integrated, which implies that
correlations between all the existing nucleons and fragments are
treated in all orders. This is especially important in the case when
one deals not only with two IMF's, but with many. In the following we
present comparisons of the QMD+SMM
fragment-fragment correlation function for the reaction
Fe (100 A~MeV) + Au for central collisions (b = 0 - 6 fm) in
to the data of Ref.~\cite{sang92}. After the break-up of
the hot target remnant within the SMM model (after
300 fm/c reaction time) we neglect the nuclear forces and follow the
Coulomb trajectories of the charged particles through an experimental
aceptence filter.

The QMD+SMM approach models the dynamical reaction and subsequent
statistical decay of the excited pre fragments (details about the two
models can be found in refs.~\cite{QMD2,sang92,smm}).  Previous
investigations with the QMD model \cite{QMD2} have shown that there
are two different mechanisms leading to the multifragmentation. One is
related to the mechanical rupture of the system whenever compressional
effects are important; the other produces fragments thermally from an
equilibrated source.  This thermal multifragmentation has so far not
been described in a microscopic model like QMD
\cite{QMD2}.  In the first comparison to inclusive IMF data  it was
shown that a two--step
model was necessary to reproduce the experimental angular
distributions \cite{sang92}.  This two--step model involved the
calculation of initial kinematics and excitation energy of all pre
fragments with the QMD model followed by a subsequent deexcitation
calculation utilizing the Statistical Multifragmentation Model (SMM)
of Botvina et al. \cite{smm}.  The input for the SMM stage of the
reaction, i.e., the mass and the excitation energy of the fragments,
are consistently determined within the QMD approach.

The SMM model describes the multifragmentation of highly excited
nuclei based on the statistical approach and a liquid--drop
description of hot fragments. It is assumed that the excited
primordial fragments break up into an assembly of nucleons and
fragments. All these decay products are described as Boltzmann
particles in a Freeze--Out volume $V = V_0 ( 1 + \kappa)$, where
$\kappa$ is a model parameter and $V_0$ is the volume of the system
corresponding to normal nuclear matter density. Since all the produced
fragments are excited (only particles with A $\le$ 4 are considered as
elementary particles) the final deexcitation is treated as a Fermi
break up for lighter fragments ( A $<$ 16) and via an evaporation of
nucleons and clusters up to $^{18}O$ for heavier fragments (for
details see Ref.~\cite{smm}).  In the following we will vary the
volume parameter $\kappa$ from 1 to 15 in order to extract the
Freeze--Out volume by comparing the reduced velocity correlation
functions for each value of $\kappa$ with the experimental data.

Figure \ref{fig1} shows a comparison between the experimental data
\cite{sang92} and the QMD+SMM calculations using $\kappa = 2,5$ and
$10$ for two different projections of the double differential cross
section, $d^2\sigma/dZ/d\Omega$. Both the data and the calculations
are for central collisions only (for details see Ref.~\cite{sang92}).
The upper part of figure \ref{fig1} shows the charge yield
distribution at a laboratory angle of $72^{\circ}$ ($\pm 13^{\circ}$),
while the lower part shows the angular distribution of fragments with
Z=10. In both cases it can be seen that the variation of the volume
parameter $\kappa$ does not influence these semi-exclusive observables
and all calculations agree reasonably well with the data. However, the
calculated angular distributions are still somewhat too
steep and under predict the data at backward angles.

The reduced velocity correlation functions shown in figure \ref{fig2}
have been obtained by taking the ratio
of the correlated reduced velocity distribution
and a background distribution obtained by event mixing
between correlated IMF's in a
single event divided by the same quantities obtained from two IMF's
from different physical events, thereby eliminating final state
correlations.  In all cases the true ($Y_{true} (v_{red})$) and
background distributions ($Y_{back} (v_{red})$) were developed
separately for two classes of events: 1) two IMF's detected on opposite
sides of the beam and, 2) two IMF's detected on the same side of the
beam.  In the areas where these two correlation functions overlap in
relative velocity it was found that the correlations obtained
were effectively identical and so in this paper
the results have, in all cases, been combined into single
distributions.  The two fragment correlation function is then
calculated according to %
\begin{equation}
\label{defcorr}
1 + R(v_{red}) = \frac{Y_{true} (v_{red})}{Y_{back} (v_{red})}.
\end{equation}

Fig.~\ref{fig2} shows the experimental correlation function (symbols)
for the three different charge product bins $Z_1 \cdot Z_2 = 25-64 \;$
, $\;65-129\;$ and $\;130-250$ (the smallest detected charge is Z=5 in
all cases) compared to the calculated results (lines) for different
volume parameters $\kappa$ in the SMM model (recall that $V = V_0
(1+\kappa)$). The curves show fits to the theoretical correlation
functions using the fitting function %
\begin{equation}
\label{fit}
1 + R(v_{red}) = a \frac{ 1 + e^{ \frac{ d-v_{red} }{e} } } {1 + e^{
\frac{ b-v_{red} }{c} } } .
\end{equation}

It can be seen that, in the limit of simultaneous fragment emission,
the size of the Coulomb hole can be explained within the QMD+SMM
approach if a volume parameter between $\kappa = 2 \;$ and $\;10$ is
used; both the results with $\kappa =1$ and $\kappa = 15$ clearly
disagree with the data for the heavier fragments.  This means that
these correlation functions are consistent with a
\underline{simultaneous} break--up with a Freeze--Out density of
$\varrho = 0.1 - 0.3 \; \varrho_0$.  For the lightest fragments the
smallest volume parameter seems to describe the data best, but here
the sensitivity to the Freeze-Out volume is less pronounced. This may
indicate that the fragments do not come from a single Freeze-Out
volume, but that the smaller fragments are emitted earlier from a smaller
volume.

These results are in good agreement with previous investigations of
the same data using the sequential three-body trajectory code MENEKA
\cite{craig}. In this case an emission time of order 500 fm/c
was found for a fixed freeze--out volume equivalent to $\kappa = 1$.
These time scales are comparable to the typical time a fragment needs
to traverse a Freeze-Out volume of roughly 5 times the nuclear volume.

Another feature that can be observed in the calculated correlation
functions in figure
\ref{fig2} is the  pronounced enhancement  at a $v_{red} \approx 15-20$. This
additional correlation is absent in the present light fragment data
and is not seen in the data of the MSU group \cite{kim} except for
very peripheral collisions \cite{bowman}. Our calculations also show
that this effect is more pronounced for larger impact parameters and
increases (see figure \ref{fig2}) with decreasing SMM volume parameter
(a similar behavior has recently been found in the Berlin model
\cite{schapiro} when the excitation energy is decreased). A similar
correlation was recently observed for the Au + Au reaction at 150 A
MeV bombarding energy \cite{ross}. In this case the enhancement has
its origin in the collective flow of the fragments.  In our case,
however we are dealing with a more asymmetric reaction at lower
energies where the collective flow vanishes due to the compensation of
the repulsive and attractive nuclear forces. However, the Coulomb repulsion
due to a heavy third fragment can
produce an additional correlation in the fragment--fragment
correlation function.  In table~\ref{table1} we show the average charge of the
largest fragment as well as the average charge asymmetry $<\Delta Z> =
\sqrt{( (Z_1 - Z_2)^2 + (Z_1 - Z_3)^2 + (Z_2 - Z_3)^2)/3} $ and the
average multiplicity of IMF's with Z=4-20 for values of $\kappa$ from
1 to 15 over the impact parameter interval 0-6 fm in the reaction Fe
(100 A MeV) + Au. It can be seen that the charge distribution gets
very asymmetric with decreasing $\kappa$ and the average IMF
multiplicity decreases by a factor of two (the same holds also when
the impact parameter is increased for a fixed volume parameter). This is
a strong indication that the enhancement in the calculated correlation
functions in figure \ref{fig2} results from a large charge asymmetry.

The fact that the experimental data in ref.~\cite{craig} does not show
such an enhanced peak indicates that the break up pattern in nature is
more symmetric than described within the SMM model. In order to
clarify this point we show in figure \ref{fig3} the experimental data
for events with three detected IMF's. For this subset of the data,
only the two lightest fragments are used to generate the correlation
functions; the heaviest fragment is simply tagged as $Z_{max}$.  The
curves in Fig. \ref{fig3} have been generated based on the following
$Z_{max}$ selection criteria: a) the numerator in eq.~\ref{defcorr}
contains only pairs from events with $Z_{max} \ge 18$ (note however
that this refers to the largest \underline{detected} fragment) (dotted
line), while no requirement has been applied to the fragments in the
background pairs (the denominator in eq.~\ref{defcorr}); b) both the
correlated and the background pairs are from events with detected
$Z_{max} \ge 18$ (dashed line); c) no cuts on $Z_{max}$ (full line).

One clearly observes that, in accordance to the QMD+SMM results, a
large peak is found for the trigger condition a) due to the recoil of
a heavy third fragment which is not accounted for in the background. This
peaks goes away when the background fragments are taken from events
with a similar charge asymmetry. The full line shows again that the
enhancement virtually disappears when no cuts are made on $Z_{max} $.


We have shown that the fragment--fragment reduced velocity correlation
functions can be explained within the two--step model QMD+SMM, where
the initial, dynamical step of the reaction, modeled using the
microscopic QMD approach, is followed by a simultaneous statistical
multifragmentation.  The correlation functions for the system Fe
(100 A MeV, b=0-6fm) + Au are consistent with a
\underline{simultaneous} break--up with a Freeze--Out density of
$\varrho = 0.1 - 0.3 \; \varrho_0$.  These findings are consistent
with previous investigations assuming a sequential emission of the
fragments from a fixed volume source. The typical time scal for
this process was found to be less than 500 fm/c. We note that these
time scales are comparable to the typical time a fragment needs to
traverse a Freeze-Out volume of roughly 5 times the nuclear volume.
The calculations show a additional correlation due to the coulomb
repulsion from a heavy third fragment in an asymmetric decay. This
effect is less pronounced in the experimental data, unless one
explicitly triggers on pairs from events with a large charge
asymmetry.

{\bf Acknowledgements:}

This work was supported by the US Department of Energy by LLNL under
Contract W-7405-ENG-48. One of us (G.P.) gratefully acknowledges
support from the Wissenschaftsausschuss of the NATO via the DAAD.


\clearpage
\noindent
{\Large {\bf Figure Captions:}}
\\[1cm]

\begin{figure}[h]
\caption{\label{fig1}}

Projections of the semi--exclusive triple differential cross section
$d^2 \sigma/dZ d\Omega$ for the reaction Fe (100 A~MeV)
+ Au.  The symbols show the data \cite{sang92} for the fragment charge
distributions at $\vartheta_{Lab} = 72^{\circ}$ (upper panel) and the
fragment angular distributions for $ Z = 10$ (lower panel). The
histograms show the calculations with the QMD+SMM (b = 0 - 6 fm)
model for different
volume parameters, $\kappa$, as indicated.

\end{figure}

\begin{figure}[h]
\caption{\label{fig2}}
Mixed fragment reduced velocity correlation functions for three
different constraints on the Coulomb product $Z_1 \cdot Z_2$. The
symbols show the data of ref.~\cite{craig}, while the curves are fits
to equation \ref{fit} obtained with the QMD+SMM model.  The different
lines show calculations done with the volume parameter $\kappa = 1$
(solid line), $\kappa = 2$ (variable length dashed line), $\kappa = 5$
(dashed line), $\kappa = 10$ (dash--dotted line) and
$\kappa = 15$ (dotted line),

 \end{figure}

\begin{figure}[h]
\caption{\label{fig3}}
Experimental correlation function for the lightest two fragments in
events with three detected IMF's. The full line shows the fit to the
correlation function obtained from all events while for the dotted
line includes only events in which the largest of the three detected
fragments has a charge greater than 17.  The dashed line shows the fit
when both the true and background pairs are from events with $Z_{max}
\ge 18$ (dashed line).

 \end{figure}

\clearpage

\noindent
{\Large {\bf Tables:}}

%
%
%
%

\begin{table}[h]
\begin{center}
\begin{tabular}[h]{c|lll}
$\kappa$ & $<Z_{max}>$ & $<\Delta Z>$ & $<MUL_{Z=4-20}> $
\\ \hline
1 & 26.5 & 12.3 & 2.2\\
2 & 22.7& 10.5 & 2.7 \\
5 & 18.8 & 8.2& 3.4 \\
10 & 17.9 & 7.5 & 3.6 \\
15 & 16.0 & 6.2 & 4.0
\end{tabular}
\end{center}
\caption { \label{table1} }
The average charge of the largest fragment as well as the average
charge asymmetry $<\Delta Z> = \sqrt{( (Z_1 - Z_2)^2 + (Z_1 - Z_3)^2 +
(Z_2 - Z_3)^2)/3} $ and the average multiplicity of IMF's with Z=4-20
for the reaction Fe (100 A MeV, b=0-6 fm) + Au and values of $\kappa$
from 1 to 15.
\end{table}
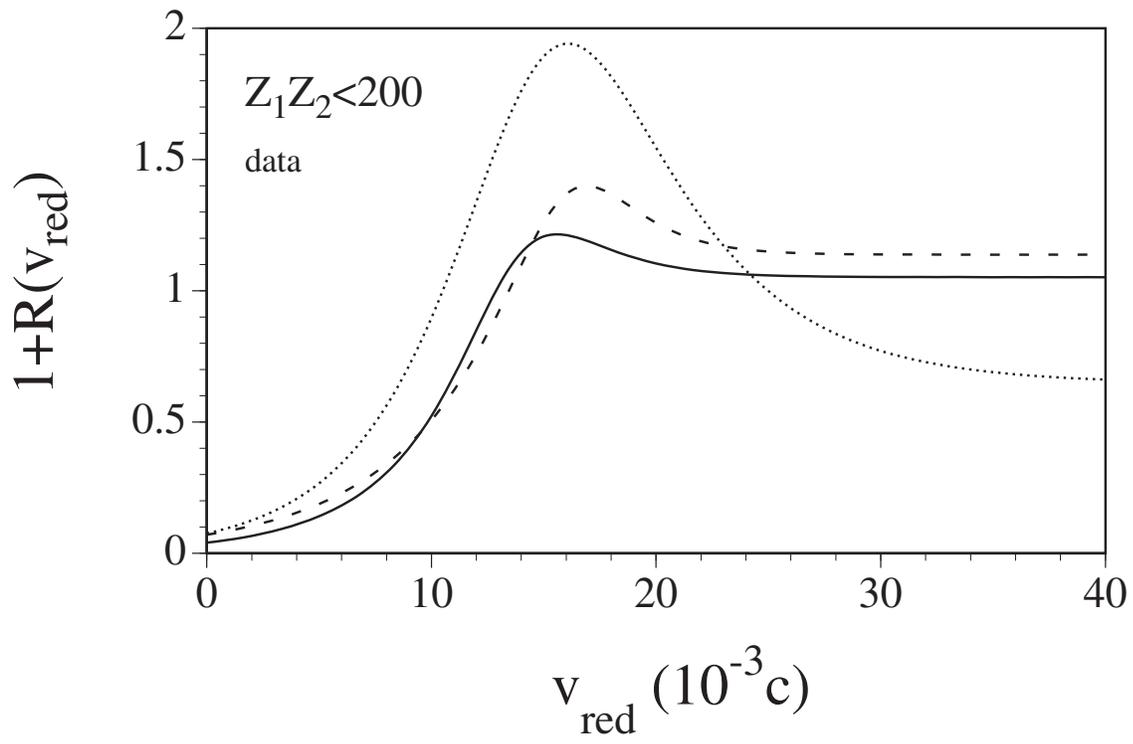
\end{document}